\documentclass[aip,prm,twocolumn,superscriptaddress,floatfix,10pt]{revtex4}
\usepackage{graphicx}
\usepackage{amsmath}
\usepackage{color}
\begin{document}

\title{Smart polymer solution and thermal conductivity: How important is an exact polymer conformation?}

\author{Mokter M. Chowdhury}
\affiliation{Quantum Matter Institute, University of British Columbia, Vancouver BC V6T 1Z4, Canada}
\author{Robinson Cortes--Huerto}
\affiliation{Max Planck Institute for Polymer Research, Ackermannweg 10, 55128 Mainz, Germany}
\author{Debashish Mukherji}
\email[]{debashish.mukherji@ubc.ca}
\affiliation{Quantum Matter Institute, University of British Columbia, Vancouver BC V6T 1Z4, Canada}

%\date{\today}

\begin{abstract}
Heat management in devices is a key to their efficiency and longevity. 
Here, thermal switches (TS) are of great importance because of their 
ability to transition between different thermal conductivity $\kappa$ states. While traditional TS are bulky and slow, recent experiments have
suggested ``smart" responsive (bio--inspired) polymers as their fast alternatives. One example is poly(N--isopropylacrylamide) (PNIPAM) in water, where $\kappa$ drops suddenly around a temperature $T_{\ell} \simeq 305$ K when a PNIPAM undergoes a coil--to--globule transition. At a first glance, this may suggest that the change in polymer conformation has a direct influence on TS. However, it may be presumptuous to trivially ``only" link conformations with TS, especially because many complex microscopic details control macroscopic conformational transition. Motivated by this, we study TS in ``smart" polymers 
using generic simulations. As the test cases, we investigate two different modes of polymer collapse using external stimuli, i.e., changing $T$ and cosolvent mole fraction $x_{\rm c}$. Collapse upon increasing $T$ shows a direct correlation between the conformation and $\kappa$ switching, while no correlation is observed in the latter case. 
These results suggest that the (co--)solvent--monomer interactions play a greater important role than the exact conformation in dictating TS. 
While some results are compared with the available experiments, possible future directions are also highlighted.
\end{abstract}

\maketitle

\section{Introduction}
\label{sec:intro}

Thermal conductivity coefficient $\kappa$ is a key physical property that dictates the use of a material for its 
wide range of applications~\cite{CHOY1977984,PolRevTT14,Keblinski20}. Here, one of the grand challenges is to attain a material with high $\kappa$,
especially when used under the high temperature $T$ conditions~\cite{Pipe15NMat,Cahill16Mac}. This is particularly important for the fast removal of 
excess heat from the devices for their efficient use and longevity. Typical examples of such devices include,
but are not limited to, thermal diodes~\cite{pnipamLCSTkappa}, nano--scale devices~\cite{Cahill11PRB,Pipe15NMat}, 
thermal bio--materials~\cite{tomko2018tunable}, thermal switching (TS)~\cite{tian18acsml}, and/or
smart batteries~\cite{AShank17ScAdv}. One of the key questions here is to attain a fast switching between different thermal 
transport coefficient $\kappa$ states~\cite{pnipamLCSTkappa,tian18acsml,thermS19jpcc}. While the traditional TS suffer from slow tunability, 
the recent experiments have suggested to use ``smart" responsive polymers.

A polymer is referred to as ``smart" responsive when a small change in the external stimuli can change its
phase by a large degree~\cite{Mukherji20AR,papadaks19lang}. One typical example is the lower critical solution temperature (LCST) behavior of the 
smart polymers~\cite{DGbook,DoiBook,DesBook}. For $T < T_{\ell}$, a polymer remains expanded due to a strong solvent--monomer interaction
and the solvent cage around it. When $T > T_{\ell}$, a certain number of solvent--monomer contacts break and 
thus a chain collapses into a globule. Here, $T_{\ell}$ is known as the LCST. 
There are many polymers that show LCST transition with varying $T_{\ell}$ that depends 
on the specific monomer structures~\cite{chi98prl,lutz06jacs,lcstPols,tiago18pccp}, i.e., $T_{\ell}$ can also be tuned via co--polymerization~\cite{koberstein16mac,TakahiroJCP2017} 
and/or tacticity~\cite{copolPNI,copol1}.

A most common smart polymer is poly(N--isopropylacrylamide) (PNIPAM) that shows an LCST behavior in water with $T_{\ell} \simeq 305$ K~\cite{chi98prl,Mukherji17JCP}.
Within the context of PNIPAM solvation in water, recent experiments have shown that the LCST--transition of PNIPAM can be used as a TS application~\cite{pnipamLCSTkappa,tian18acsml}. 
In particular, one experimental study had shown that $\kappa$ of PNIPAM--water solution follows the same trend as its LCST--transition, 
i.e., $\kappa$ changes from about 0.61 W/Km (for $T < T_{\ell}$) to about 0.53 W/Km (for $T > T_{\ell}$), with a transition around
$T_{\ell} \simeq 305$ K~\cite{tian18acsml}. 
These results were obtained for the PNIPAM concentrations $c$ below their ``so called" chain overlap concentration $c^*$~\cite{DGbook,DoiBook,DesBook},
i.e., in the dilute (single) chain limit. 
%This is also why the TS signal becomes more prominant for the higher $c$ values in Ref.~\cite{tian18acsml} even when $c<c^*$. 
Note that we have estimated $c^*$ from the molecular weights used in these experiments $M_{\rm w} \simeq 3 \times 10^4$ g/mol~\cite{tian18acsml}. 
Another set of experiments on a PNIPAM--PDMS diode revealed an opposite trend, i.e., the thermal resistance $R$ shows a sudden drop 
(or an increase in $\kappa$) around $T \simeq T_{\ell} =305$ K~\cite{pnipamLCSTkappa}. The latter study used $c > c^*$. 

The experimental results clearly highlight 
the importance of coil--to--globule transition in dictating TS in the polymer solutions. 
However, given that an LCST transition is dictated by a delicate balance between the water structure within the
first solvation shell around a PNIPAM (i.e., within 0.5 nm from the PNIPAM backbone), 
interaction/ordering between the water molecules, and different conformational/compositional fluctuations across 
$T_{\ell}$~\cite{DMKKpolRev23}, it is rather presumptuous to trivially ``only" link $\kappa$ and the conformational changes. 

Motivated by the above discussion, we employ generic molecular simulation to study the TS of smart polymers that 
to the best our knowledge is lacking. In particular, to decouple the effect of exact conformational states and other 
microscopic details, we investigate two different modes of polymer collapse. In one case we model
the above discussed experimental system with changing $T$~\cite{pnipamLCSTkappa,tian18acsml}, while the
second case uses smart polymer collapse in a mixture of two good solvents at a given $T$, 
aka the co--nonsolvency (CNS) phenomenon~\cite{muthukumar91mac,winnik11mac,mukherji13mac}.
In a nutshell, CNS is associated with a polymer collapse in a mixture of two individual good and
fairly miscible solvents. These two modes of collapse have different microscopic 
coordination and thus provide ideal cases for our study. 

The remainder of this manuscript is organized as follows: in Section~\ref{sec:method}, we sketch the key ingredients of
the method and model used in this study. The results are highlighted in Section~\ref{sec:res} and finally the conclusions are drawn in Section~\ref{sec:conc}.

\section{Model and method}
\label{sec:method}

\subsection{Basic model}

A chain is represented by the well--known bead--spring model of polymers~\cite{kremer1990dynamics}. 
Within this model, two bonded monomers interact via a combination of repulsive Lennard--Jones (LJ) potential with a 
cut--off distance $r_{\rm c} = 2^{1/6}\sigma$ and the finite extensible nonlinear elastic (FENE) potential. 
The monomer size is taken as $\sigma_{\rm m} = 1.0\sigma$ and the LJ interaction energy 
$\epsilon_{\rm bond} = 1\epsilon$. The effective bond length of this model is $\ell_{\rm b} \simeq 0.97\sigma$. 
The results are presented in the units of LJ energy $\epsilon$, LJ length $\sigma$, mass of a
monomer $m$, time $\tau = \sigma\sqrt{m/\epsilon}$, pressure $p_{\circ} = \epsilon/\sigma^3$, and leads to a generic unit of thermal conductivity $\kappa_{\circ} = k_{\rm B}\tau/\sigma$. $k_{\rm B}$ is the Boltzmann constant whose value is unity in a generic model~\cite{kremer1990dynamics}

Two different modes of polymer collapse are modelled by tuning the interactions between
different solution components that can mimic--
(1) LCST--like transition: $T-$dependent 
monomer--monomer and monomer--solvent interactions.
(2) CNS transition: additionally to the parameters in part (1) at a given $T$,
monomer--cosolvent, solvent--cosolvent, and cosolvent--cosolvent interactions.
These details will be discussed in the results Section~\ref{sec:res}.

\subsection{Molecular dynamics simulations}

The degree of polymerization $N_{\ell} = 50$ is chosen for a polymer chain, where one monomer bead
corresponds to one Kuhn length $\ell_{\rm k}$~\cite{kremer1990dynamics}. Polymer chains are solvated in $N = 5 \times 10^4$ 
solvent molecules. In the case of polymer solvation in binary mixtures, the cosolvent mole fraction 
is varied between $0.0 \le x_{\rm c} \le 1.0$, i.e., $x_{\rm c} = 0.0$ and 1.0 represent the pure solvent and
the pure cosolvent phases, respectively. In some cases, multiple chain systems are also simulated,
i.e., the polymer concentrations $c \to 0$ (single chain limit), $c \simeq c^*$, and $c > c^*$ (semi--dilute regime). 
Details will be discussed whenever appropriate within the draft.

Simulations in this study are employed using the LAMMPS molecular dynamics package~\cite{lammps}.
The equations of motion are integrated using the velocity Verlet algorithm with a time step $\Delta t = 5 \times 10^{-3} \tau$.
The temperature is varied between $T = 0.90-1.20\epsilon/k_{\rm B}$, which is imposed using the Langevin thermostat 
with a damping coefficient of $\gamma = 0.5\tau^{-1}$. Individual configurations are first equilibrated under 
a constant pressure $p$ ensemble for $10^7$ MD steps, with $p = 10 \epsilon/\sigma^2$.
%The pressure coupling is employed using a Nos\'e Hoover  
The production runs are performed for an additional $10^7$ MD steps under the canonical ensemble. 
This choice is over two orders of magnitude longer than the typical relaxation time for $N_{\ell} = 50$. 
During the production runs observables, such as the single chain gyration radius $R_{\rm g}$ and
structure factor $S(k)$ are calculated. 

\subsection{Thermal transport coefficient}

The final configurations from the above runs are then used for $\kappa$ calculations using the 
equilibrium Kubo--Green method~\cite{ZwanzigRev} implementation in LAMMPS \cite{lammps}. 
For this purpose, the equations of motion are integrated in the microcanonical ensemble.
The heat flux autocorrelation function $C(t) = \langle {\bf J}(t) \cdot {\bf J}(0) \rangle$ 
is obtained by sampling the heat flux vector ${\bf J}(t)$. %The typical $C(t)$ data is shown in the Supplementary Fig. S1 \cite{epaps}.
Here, $C(t)$ is sampled over a time of $0 \leq\ t \leq 50 {\tau}$ with $\triangle t = 0.005{\tau}$, 
which is one order of magnitude larger than the typical de-correlation time of $0.5-1.0\tau$. %, see the Supplementary Fig. S1 \cite{epaps}.
The correlation data are accumulated over $t_{\rm c} = 5 \times 10^4 {\tau}$.
Finally, $\kappa$ values are calculated by taking the plateau of their cumulative integrals,
\begin{equation}
        \kappa(t) = \frac{v}{3k_{{\rm B}}T^{2}}\int_0^{\mathcal{T}} C(t) {\rm d}{t},
        \label{eq:kgint}
\end{equation}
where $v$ is the system volume. Ideally the sampling time $\mathcal{T} \to \infty$.
A few examples of $\kappa(t)$ are shown in the Supplementary Fig.~S1.
$\kappa$ is calculated by taking an average between $20\tau \le \mathcal{T} \le 40\tau$ from
the plateau of $\kappa(t)$. 
Individual data sets are average of twenty statistically independent runs and the error bars are their standard deviations.

\section{Results}
\label{sec:res}

\subsection{$\kappa$ under a LCST--like transition}

\subsubsection{Component--wise interactions}
\label{ssec:lcstInt}

We first consider the model polymers in a single component solvent that can mimic
an LCST--like transition. Note also that we only call this a ``LCST--like" because we include $T$ effects 
via the $T-$dependent 2--body LJ interaction parameters. Here, a key for such a model is to properly 
describe the solvent--solvent, solvent--monomer, and monomer--monomer interactions. 

The size of a solvent molecule is taken as half of that of a monomer $\sigma_{\rm s} = \sigma_{\rm m}/2$. This choice is motivated by the solvent--monomer
size asymmetry known from an all--atom system of PNIPAM in water~\cite{Mukherji17JCP}.
The solvent molecules also interact using an attractive LJ interaction with $r_{\rm c} = 2.5\sigma_{\rm s}$. 
The interaction energy is chosen as $\epsilon_{\rm s} = 0.8\epsilon$ that ensures a fluid phase over 
the range of $T$ investigated herein. The equilibrium number density of the solvent molecules is 
$\rho \simeq 5.7 \sigma^{-3}$ at its reference pressure $p = 10p_{\circ}$ and $T = 1.0\epsilon/k_{\rm B}$.

%\subsubsection{Monomer--solvent and monomer--monomer interactions}

In our model, we have kept $\epsilon_{\rm s}$ constant at all $T$, but the non--bonded monomer--monomer $\epsilon_{\rm mm}(T)$ and 
monomer--solvent $\epsilon_{\rm ms}(T)$ energies are taken as $T-$dependent, see the Supplementary Section S2.
% ,
% \begin{equation}
% \label{eq:chiT}
% \epsilon_{\rm ij} (T) = c_1 \left[ c_2 - \frac {\left(c_3 + \frac {k_{\rm B}T} {\epsilon}\right)}{\{1 + \exp{\left(\frac{T_{\ell} - T}{\Delta T}\right)}\}} \right],
% \end{equation}
% in the LJ potential
% % \begin{equation}
% % \label{eq:LJnb}
% % v_{\rm ij}(r,T) = 4\epsilon_{\rm ij} (T) \left[ \left(\frac {\sigma_{\rm ij}}{r}\right)^{12} -\left(\frac {\sigma_{\rm ij}}{r}\right)^6 \right].
% % \end{equation}
% with $\sigma_{\rm ij} = \left(\sigma_{\rm i} + \sigma_{\rm j}\right)/2$ and $r_{\rm c} = 2.5 \sigma_{\rm ij}$. 
%
%The parameters $c_i$ and $\triangle T$ in Eq.~\ref{eq:chiT} 
These are chosen such that the typical LCST--like transition can be observed.
For this purpose, we take motivation from the all--atom simulations of PNIPAM in water~\cite{Mukherji17JCP}.
Here, a coil configuration of PNIPAM for $T < T_{\ell}$ is stabilized by 
the solvent--monomer hydrogen bonds (H--bonds) that creates a cage around a PNIPAM. 
For $T > T_{\ell}$, a certain number of H--bonds break, as a result water molecules get expelled from 
within the first solvation shell and thus gain translational entropy that is larger than the 
conformational entropy loss upon collapse. This makes LCST an entropy driven process~\cite{DGbook,DoiBook,DesBook}.
%Such a caging effect is because 
%%%%%%%%%%%%%%%%%%%%%This is a result of the complex multi--body effects that 
%originate because of a 
%delicate balance between the . 
%This can however 
%%%%%%%%%%%%%%%%%%%%%can be modelled in a 2--body LJ picture by using the $T-$dependent $\epsilon_{\rm mm}(T)$ and $\epsilon_{\rm ms}(T)$ energies.
%interaction parameter.

It is known from the all--atom simulations that the PNIPAM--water H--bonds show a sharp drop across $T_{\ell}$, while there
is only a very weak variation in NIPAM--NIPAM H--bond even when their coordination increases upon collapse~\cite{Mukherji17JCP}.
Given this all--atom data, we choose $c_1 = 0.1\epsilon$, $c_2 = 10.0$, $c_3 = 6.0$, and $\triangle T = 0.01 \epsilon/k_{\rm B}$
for $\epsilon_{\rm mm}(T)$ that shows a weak variation with $T$, see the red data set in the Supplementary Fig.~S2(a).
To mimic a sharp drop in $\epsilon_{\rm ms}(T)$, $c_1 = 0.1\epsilon$, $c_2 = 9.0$, $c_3 = 0.0$, and $\triangle T = 0.05 \epsilon/k_{\rm B}$,
see the black data set in the Supplementary Fig.~S2(a). This model has an effective $T_{\ell} \simeq 1.05\epsilon/k_{\rm B}$.
The corresponding $R_{\rm g}(T)$ is shown in the Supplementary Fig.~S1(b). 
It can be appreciated that the generic trend of an LCST--transition is reasonably captured using our model.

\subsubsection{Thermal conductivity}

\begin{figure}[ptb]
\includegraphics[width=0.49\textwidth,angle=0]{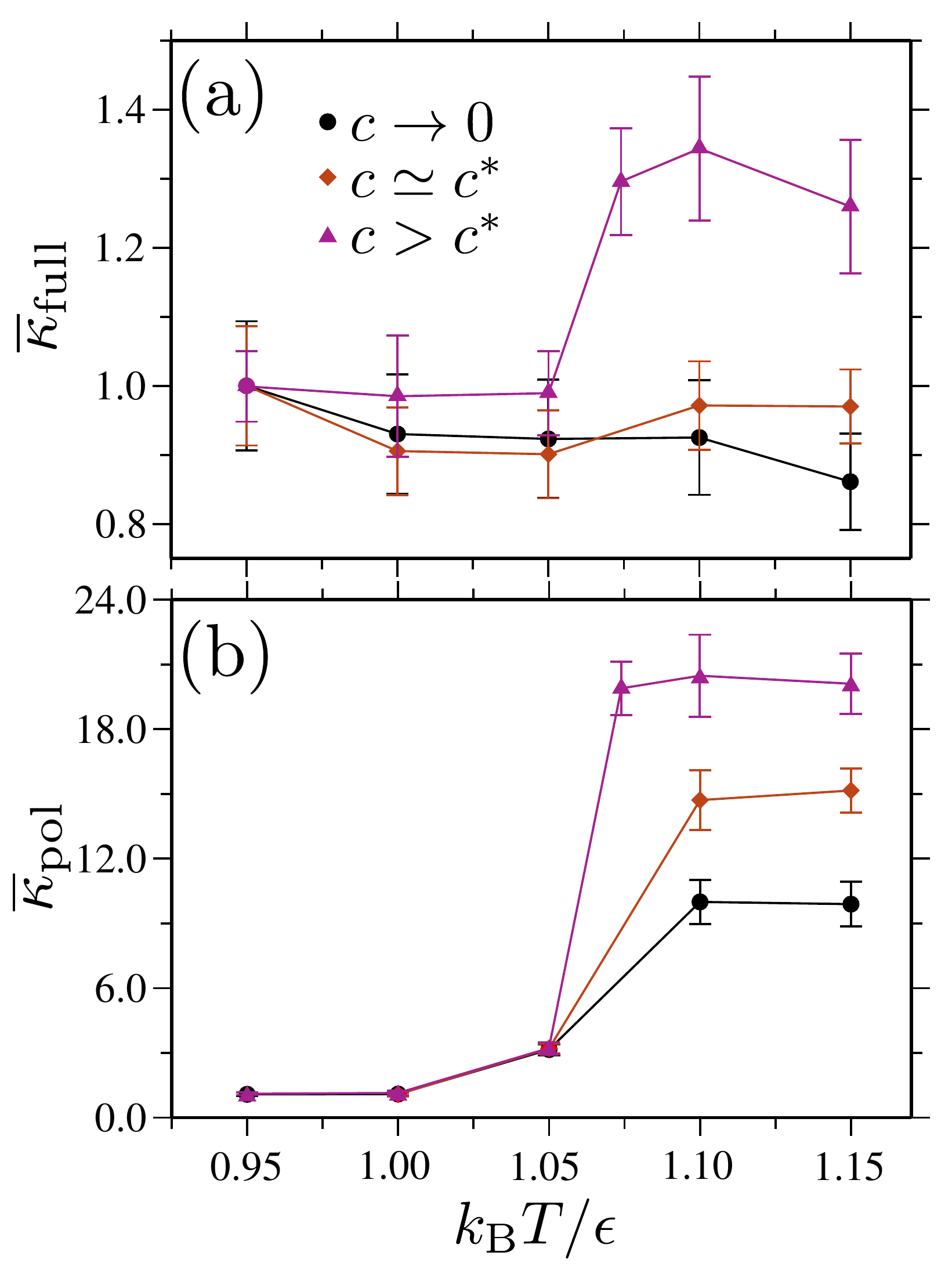}
	\caption{Parts(a--b) show the normalized thermal transport coefficient as a function of temperature $T$ 
 for the full system $\overline{\kappa}_{\rm full}$ and for only polymer (without considering solvent molecules in Eq.~\ref{eq:kgint}) $\overline{\kappa}_{\rm pol}$, respectively.
Both $\kappa$ are normalized by their values at $T = 0.95k_{\rm B}/\epsilon$, see the Supplementary Tables S1--S3. 
%Part (c) shows a ratio of the non--bonded to bonded contribution to $\kappa_{\rm pol}$. 
Data is shown for different polymer concentrations $c$.
\label{fig:kappaSol}}
\end{figure}

Fig.~\ref{fig:kappaSol}(a) shows the thermal transport coefficient of the full system $\kappa_{\rm full}$ 
as a function of $T$ for three different polymer concentrations, namely: (I) under infinite dilution $c \to 0$,
(II) $c \simeq c^*$, and (III) the semi--dilute regime $c > c^*$.\\

\begin{figure}[ptb]
\includegraphics[width=0.49\textwidth,angle=0]{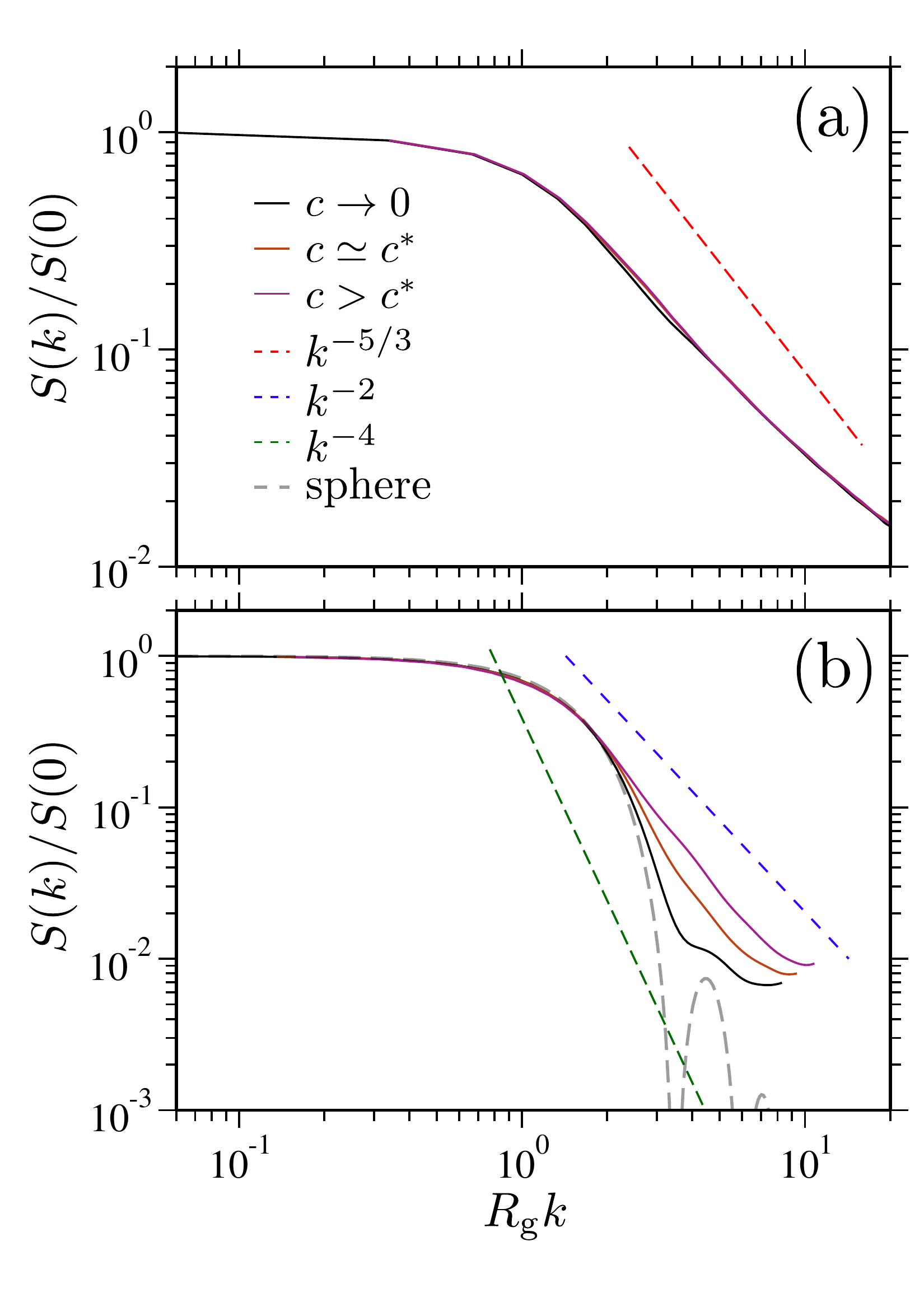}
	\caption{Single chain structure factor $S(k)$ for three different polymer concentrations $c$. $S(k)$ is normalized 
 by the chain length $S(0) = N_{\ell} = 50$ and the $x-$axis is multiplied by the gyration radius $R_{\rm g}$. 
 The representative structural scaling laws are also shown. Parts(a-b) show the data for $T = 1\epsilon/k_{\rm B}$
 and $T = 1.1\epsilon/k_{\rm B}$, respectively. The analytical form of the sphere scattering function is also drawn in part (b).
\label{fig:sofk}}
\end{figure}

\noindent{\bf (I) Infinite dilution limit ($c\to 0$):}
It can be appreciated that $\kappa_{\rm full}$ shows a weak decay with $T$ where a maximum decrease of about 10--15\% is 
observed for $T > T_{\ell}$, see the $\bullet$ data set in Fig.~\ref{fig:kappaSol}(a).
Previous experiments on PNIPAM in water have also shown a weak reduction in $\kappa_{\rm full}$ of about 10--15\% across $T_{\ell}$, 
so long as the single chain limit ($c < c^*$) holds~\cite{tian18acsml}.

In a polymer solution, $\kappa$ has the contributions from different factors-- (a) energy transfer between 
the non--bonded contacts of different solution components and between the bonded monomers, 
(b) motion of the particles in a solution, and (c) the structural fluctuations in a polymer chain. Here, the non--bonded contacts can either be van der Waals (vdW) or H--bonded,
the strength of which is 1--4$k_{\rm B}T$ under ambient temperature~\cite{desiraju02,Mukherji20AR}. 
These contacts are usually soft and thus typically have weaker energy transfer rates, while a monomer--monomer 
bond is rather strong, i.e., about 80$k_{\rm B}T$ for carbon--carbon covalent bond, and thus transfer 
energy at a faster rate~\cite{MM21acsn,MM21mac}.
%%
%In a fully stretched chain, 

Bonded contribution to $\kappa$ is most dominated in a stretched chain and usually has a high value~\cite{shen2010polyethylene,bhardwaj2021thermal}.
Moreover, every time there appears a bend or a kink along a chain, they act as the scattering/defect points for energy transfer 
and as a result $\kappa$ reduces~\cite{chen19jap}.
In our case, when $T < T_{\ell}$, a chain shows a well defined self avoiding random walk configuration~\cite{DGbook,DoiBook,DesBook}, 
represented by $S(k) \propto k ^{-5/3}$ in Fig.~\ref{fig:sofk}(a). In such a good solvent chain with larger
conformational entropy (fluctuations), energy transfer is rather weak along a chain and thus $\kappa_{\rm full}$
gets mostly dominated by the energy transfer between the monomer--solvent and solvent--solvent contacts. 

For $T > T_{\ell}$, because of the broken monomer--solvent contacts a chain collapses into a globule, as represented by the sphere scattering function $S(k) \propto k^{-4}$
in the black data set of Fig.~\ref{fig:sofk}(b). Such a globule creates interfaces (or defects within a solution) within the solution and thus acts as resistance for the heat flow. This contributes to an additional reduction in the macroscopic $\kappa_{\rm full}$. 

In summary for the $c \to 0$ regime, it is not the exact change in the chain conformation that directly influences the heat flow, rather the modification in the microscopic coordination and weaker non--bonded contacts that dominate $\kappa_{\rm full}$. Something that supports this claim is that the contribution of only polymer $\kappa_{\rm pol}$ to $\kappa_{\rm full}$ shows an opposite trend and suddenly increases for $T > T_{\ell}$, see the $\bullet$ data set in Fig.~\ref{fig:kappaSol}(b). Note that $\kappa_{\rm pol}$ is computed without considering the solvent molecules in its calculation in Eq.~\ref{eq:kgint}.

\noindent{\bf (II) $c \simeq c^*$ limit:} When the chains start to overlap in a solution (i.e., for $c \simeq c^*$), $\kappa_{\rm full}$ shows a weak non--monotonic variation with $T$, see the orange $\diamond$ data set in Fig.~\ref{fig:kappaSol}(a). The initial decrease in $\kappa_{\rm full}$ for $T <T_{\ell}$ is due to the increased
conformational fluctuations (or increased scattering) with $T$, while the chains still remain in the coil state.  
The re--entrant increase for $T > T_{\ell}$ is because of a slightly expanded chain structure in comparison to a single chain globule, as shown by the deviation from $S(k) \propto k^{-4}$ to a $k^{-2}$ scaling in Fig.~\ref{fig:sofk}(b).

\noindent{\bf (III) Semi--dilute regime ($c > c^*$):} When $c$ increases even further to $c > c^*$,
a clear switching from a low $\kappa_{\rm full}$ (for $T < T_{\ell}$) to about 30\% increase (for $T > T_{\ell}$) 
is observed, see the $\triangle$ data set in Fig.~\ref{fig:kappaSol}(a). This opposite trend from 
the $c \to 0$ regime is purely due to $c$ and is consistent with the experimental data for the semi--dilute regime~\cite{pnipamLCSTkappa}. Two factors contribute to this apparent increase that compensates for the loss in the interfacial polymer--solvent contacts--
(1) A chain in an aggregate of multiple chains (for $T > T_{\ell}$) does not collapse to a globule, rather remains expanded due to the neighboring chain crowding, as represented by $S(k) \propto k^{-2}$ in Fig.~\ref{fig:sofk}(b). In such an expanded chain, bonded energy transfer can contribute significantly.
(2) A chain in such an aggregate has smaller conformational fluctuations due to crowding.
Collectively, the above two microscopic factors influence
the overall increase in $\kappa_{\rm full}$ and $\kappa_{\rm pol}$. Something that supports this claim is that the bonded and the non--bonded contributions to $\kappa_{\rm pol}$ increase by over an order of magnitude across $T_{\ell}$, see the Supplementary Table~S3.

We also note in passing that the change in non--bonded contribution to $\kappa_{\rm pol}$ is always 
about 20\% higher than the corresponding bonded contributions for $T > T_{\ell}$. 
It is because of the increased non--bonded contacts upon collapse/aggregation. 

The three bullet points discussed above further suggest that a delicate balance between inter-- and intra--chain 
coordination and interactions between the solution components control macroscopic $\kappa_{\rm full}$
at a given $c$ regime. On the contrary, the exact chain conformation may or may not play a dominant role.
Therefore, to further decouple the effect of exact conformation from the interactions, we will now investigate another mode 
of polymer collapse and its influence on $\kappa_{\rm full}$.

\subsection{$\kappa$ under the co--nonsolvency transition}

The second mode of polymer collapse investigated in this study is CNS. To model the generic features of CNS, we have used a microscopic picture proposed by one of us in Refs.~\cite{mukherji13mac,Mukherji14NC}. 
In a nutshell, this picture suggested the preferential binding of the cosolvent molecules with the monomers as a driving force for the polymer collapse. Furthermore, the free energy contrast between solvent--monomer and cosolvent--monomer was
found to be $\Delta {\mathcal F} \simeq 4k_{\rm B}T$ for PNIPAM in aqueous alcohol mixtures~\cite{mukherji13mac}.

One of the prerequisites here is that the bulk solvent--cosolvent mixtures remain fairly miscible over the full range of $x_{\rm c}$~\cite{Mukherji14NC,mukherji15jcp}.
For this purpose, our prior studies have modelled the bulk solutions using a purely repulsive LJ potential~\cite{Mukherji14NC,Zhao:2020}. 
The relative particle interactions and sizes were tuned such that the solvent--cosolvent Flory--Huggins interaction parameter $\chi \simeq 0$.
Moreover, for our present study, we have re-parameterized a model that can reproduce the generic structural and $\kappa-$behavior of the water--ethanol mixtures, while keeping $\chi \simeq 0$. One key change that we have introduced here is the attractive energy between the particles, which is important within the context of $\kappa$ behavior.

Note also that the goal of our present study is to study possible TS in solutions, we only sketch some key details of CNS herein
that are important for the present study. A general overview can also be found in Refs.~\cite{Mukherji20AR,DMKKpolRev23}.

\subsubsection{Component--wise interactions}

For the CNS transition, we have taken a reference of polymer in aqueous ethanol mixtures~\cite{mukherji17acsml,Zhao:2020}. The solvent--solvent, solvent--monomer, and monomer--monomer interactions are taken to be the same as in Section~\ref{ssec:lcstInt}. Note that the values of $\epsilon_{\rm ms}(T)$ and $\epsilon_{\rm mm}(T)$ are taken for $T = 1.0\epsilon/k_{\rm B}$. 

To mimic the relative size of an ethanol molecule with respect to a water and a NIPAM, the size of a cosolvent molecule is taken as $\sigma_{\rm c} = 0.9\sigma$~\cite{Zhao:2020}. As a first step, the solvent--cosolvent $\epsilon_{\rm sc}$ is tuned such that the two key mixture properties are reproduced: \\
(a) Our generic model shows the representative dip in the total number density $\rho$ with $x_{\rm c}$ for the water--ethanol mixtures, known from the all--atom calculations~\cite{Mukherji18JPCM} and in the experiments~\cite{perera06jcp}, see the Supplementary Fig.~S3(a). 
% \begin{table}[ptb]
% \caption{A table listing the Lennard--Jones interaction parameters used to mimic the co--nonsolvency (CNS) transition. The solvent--monomer and monomer--monomer interactions are taken from Eq.~\ref{eq:chiT} at a temperature $T = 1.0\epsilon/k_{\rm B}$.}
% \begin{center}
%         \begin{tabular}{|c|c|c|c|c|c|c|c|c|c|c|c|}
% \hline
%  %%%& & & & & & &\\
%      & ~$\epsilon_{\rm ij}~[\epsilon]$~ & ~$\sigma_{\rm ij}~[\sigma]$~ & ~$r_{\rm c}~[\sigma]$~ \\
% \hline
% \hline
% \hline
% Solvent--solvent      & 0.8      &  0.5  & 1.25   \\
% Solvent--cosolvent    & 0.9      &  0.7  & 1.75   \\
% Cosolvent--cosolvent  & 0.6      &  0.9  & 2.25   \\
% %Monomer--solvent      & 0.99532  &  0.75 & 1.875  \\
% Monomer--cosolvent    & 3.0      &  0.95 & 2.375  \\
% %Monomer--cosolvent    & 3.5$^{2}$      &  0.95 & 2.375  \\
% %Monomer--monomer      & 0.87311  &  1.0  & 2.5    \\
% \hline 
% \end{tabular}  \label{tab:cns}
% \end{center}
% \end{table}

\begin{figure}[ptb]
\includegraphics[width=0.49\textwidth,angle=0]{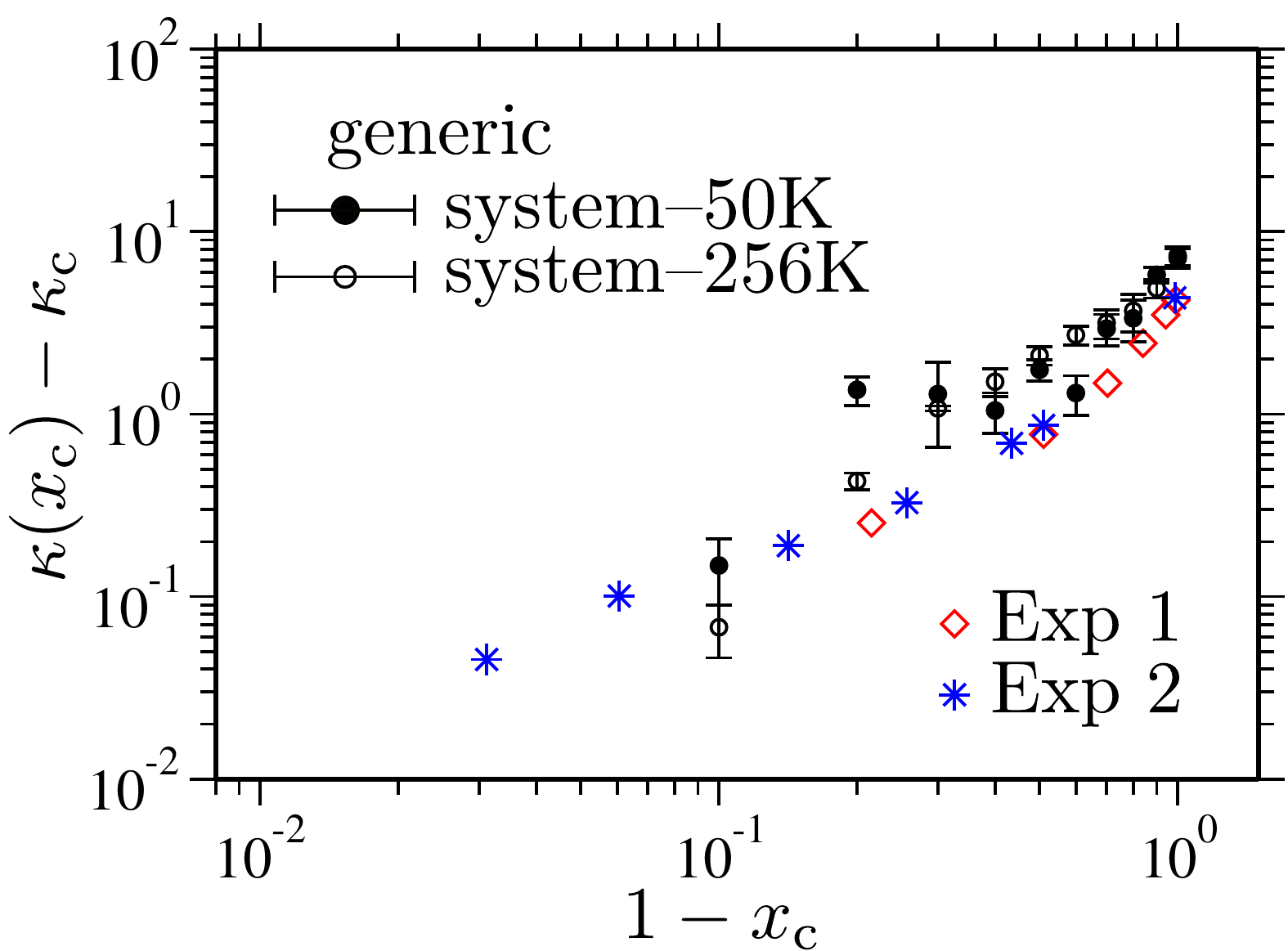}
	\caption{Scaled thermal transport coefficient $\kappa(x_{\rm c})-\kappa_{\rm c}$ of a solvent--cosolvent mixture as a function of the scaled cosolvent mole fraction $1-x_{\rm c}$. The data is shown for two different system sizes. Here, $\kappa_{\rm c} = \kappa(x_{\rm c} = 1)$, i.e., of pure cosolvent. For 50K system,  $\kappa_{\rm c} = 11.43k_{\rm B}\tau/\sigma$ and for 256K system $\kappa_{\rm c} = 11.42k_{\rm B}\tau/\sigma$. For comparison, the experimental data is also included~\cite{kappaWE}.
\label{fig:scXc}}
\end{figure}

\noindent(b) A more important property is to reproduce the trends in $\kappa$ of the bulk mixture with $x_{\rm c}$. In Fig.~\ref{fig:scXc}, it can be appreciated that our parameterization of the solvent--cosolvent interactions can reasonably capture the $\kappa$ trend reported in the experiments~\cite{kappaWE}. We note in passing that the trend in Fig.~\ref{fig:scXc} could not reproduced by using the pure repulsive interactions, hence attractive interaction is introduced.

We would also like to highlight that the two system sizes are tested because the mid--to--small systems always 
give wrong fluctuations~\cite{mukherji12jctc,robin21arx} and thus lead to unavoidable simulation artifacts. 
Since both systems in Fig.~\ref{fig:scXc} show consistent data, we have used $N = 5 \times 10^4$ for our study for the sake of computational efficiency.

Lastly, the cosolvent--monomer interaction is adjusted so that the generic coil--globule--coil transition of CNS is reproduced, see the Supplementary Fig.~S3(b) and the
representative LJ parameters in the Supplementary Table. %~\ref{tab:cns}. 

\subsubsection{Thermal conductivity}

\begin{figure}[ptb]
    \includegraphics[width=0.49\textwidth]{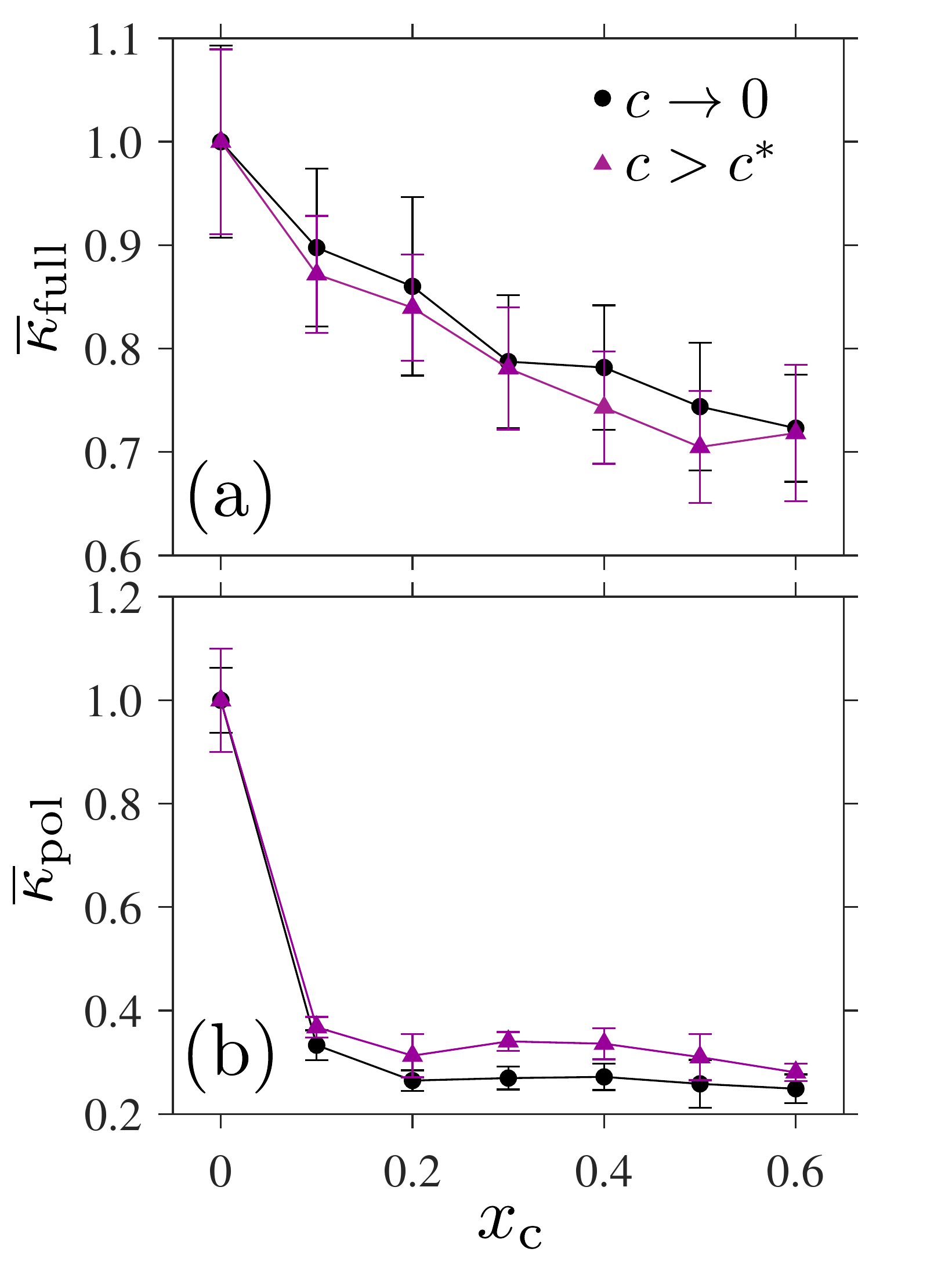}
    \caption{Same as Fig.~\ref{fig:kappaSol}, however, as a function of cosolvent mole fraction $x_{\rm c}$. The data sets are normalized with respect to their respective values for $x_{\rm c} = 0.0$, see the Supplementary Table S4.        
\label{fig:cnsXc}}
\end{figure}

\begin{figure}[ptb]
\includegraphics[width=0.49\textwidth,angle=0]{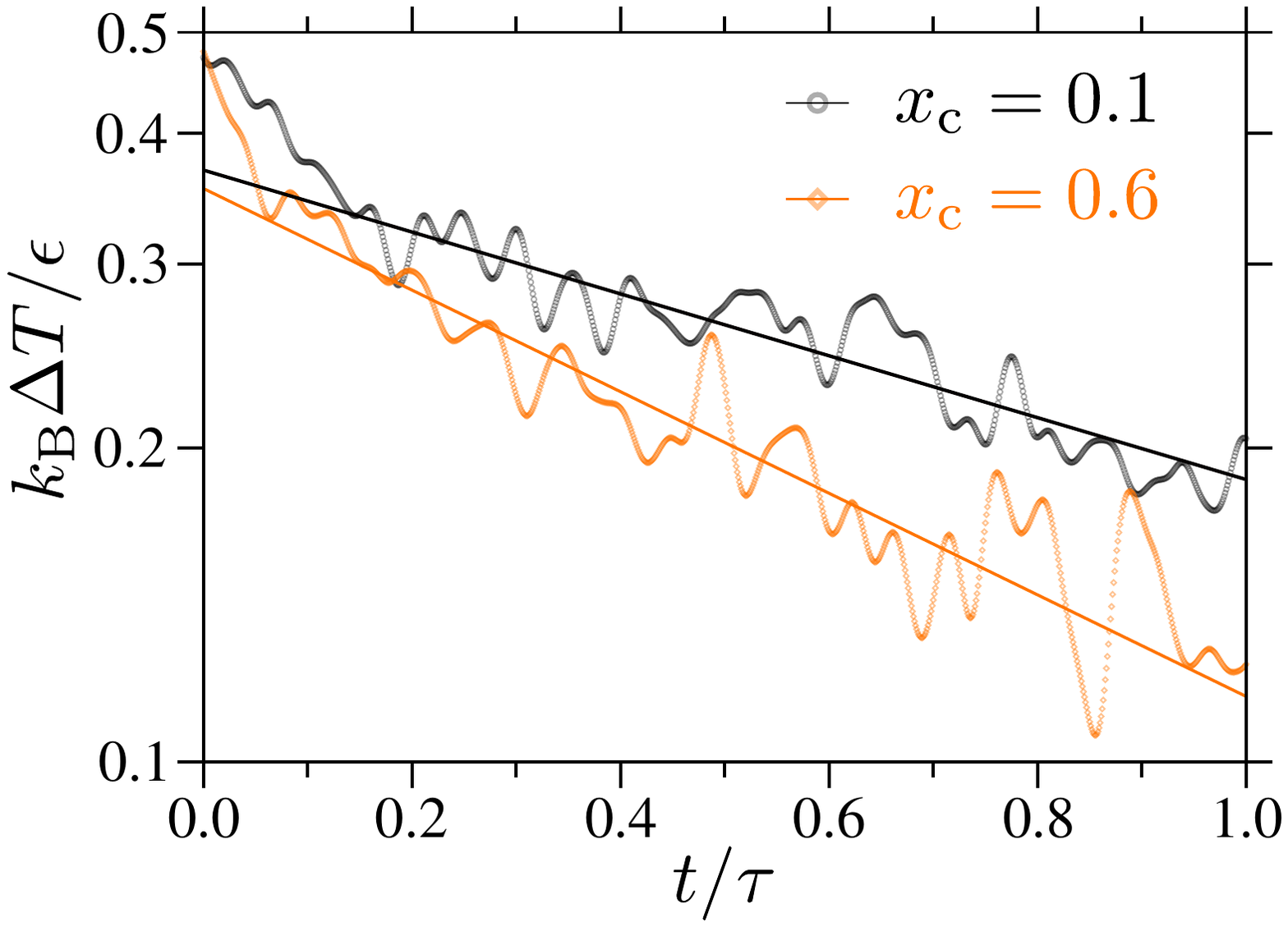}
	\caption{Relative temperature $\triangle T$ relaxation of a polymer for two different
 cosolvent mole fractions. Data is shown for a fully collapsed globule at $x_{\rm c} = 0.1$ and for an expanded chain at $x_{\rm c} = 0.6$, see the Supplementary Fig.~S3(b).
 Lines are the exponential fits $\triangle T \propto \exp(-\alpha_{\rm c} t)$ to the data with a relaxation rate $\alpha_{\rm c} = 0.68\tau^{-1}$ and $1.12\tau^{-1}$ for $x_{\rm c} = 0.1$ and 0.6, respectively.
 Note that the characteristic oscillation in Fig.~\ref{fig:Tdecay} has a frequency $\nu \simeq 20 \tau^{-1}$ that originates because of the structural fluctuations. These vibrations are stiffer than the typical FENE bond with $\nu \simeq 7\tau^{-1}$.
\label{fig:Tdecay}}
\end{figure}

Fig.~\ref{fig:cnsXc}(a) shows $\kappa_{\rm full}$ as a function of $x_{\rm c}$. A monotonic decrease with increasing $x_{\rm c}$ is clearly visible for $c \to 0$ and $c > c^*$ regimes. Furthermore, this behavior is typically the same variation as the bulk solution in Fig.~\ref{fig:scXc}. 
Note that the $x-$axis in Fig.~\ref{fig:scXc} is taken as $1-x_{\rm c}$ because this is how the experimental data was presented~\cite{kappaWE}.
Fig.~\ref{fig:cnsXc} also indicate that the variation in $\kappa_{\rm full}$ with $x_{\rm c}$ is decoupled 
from the chain conformation, see the Supplementary Fig.~S3(b). 
On the contrary, $\kappa_{\rm pol}$ in Fig.~\ref{fig:cnsXc}(b) shows a sharp drop between $0.0 < x_{\rm c} < 0.1$, which occurs at the onset of coil--to--globule transition, see the Supplementary Fig.~S3(b). Moreover, $\kappa_{\rm pol}$ remains
rather constant for $x_{\rm c} \ge 0.1$.

What causes such a qualitatively different trend in $\kappa_{\rm full}$? Why the exact chain conformations do not play a direct role under CNS?
To understand these questions, it is important to look into the influence of solvent and cosolvent molecules within the first solvation shell of a polymer in its collapsed state. 
For example, comparing the fully collapsed structures (under $c \to 0$) between an LCST collapse and a CNS collapse 
reveal that the latter is about 30\% more swollen than the former, i.e., $R_{\rm g} \simeq 1.94\sigma$ at $T = 1.15\epsilon/k_{\rm B}$ 
(see the Supplementary Fig.~S2) and $R_{\rm g} \simeq 2.9\sigma$ at $x_{\rm c} = 0.1$ (see the Supplementary Fig.~S3). 

A major difference between these two modes of collapse is that-- when a polymer collapses under LCST, 
it expels solvent molecules from within its globule structure. 
In this process, the direct monomer--monomer contacts increase significantly~\cite{chi98prl,tiago18pccp}. 
On the contrary, a CNS--based collapse is driven by the preferential binding of cosolvents with a polymer. 
When such a polymer collapses, it encapsulates a significant amount of cosolvent molecules within its globular structure
and thus leads to a more swollen globule~\cite{mukherji13mac,Mukherji14NC}. 
This solvent encapsulation was observed in the simulations of PNIPAM solvation in aqueous--alcohol mixtures~\cite{mukherji13mac,Mukherji14NC}, in proton nuclear magnetic resonance measurements~\cite{mukherji16SM}, and in the viscosity measurements~\cite{backes17lang}.
Note that the definition of solvent encapsulation (or sticky contacts) is defined by the resident time $t_{\rm r}$ of a cosolvent on a monomer. When $t_{\rm r}$ is about an order 5--10 times larger than the chain diffusion time, a cosolvent is considered to be encapsulated.

\begin{figure}[ptb]
    \includegraphics[width=0.49\textwidth]{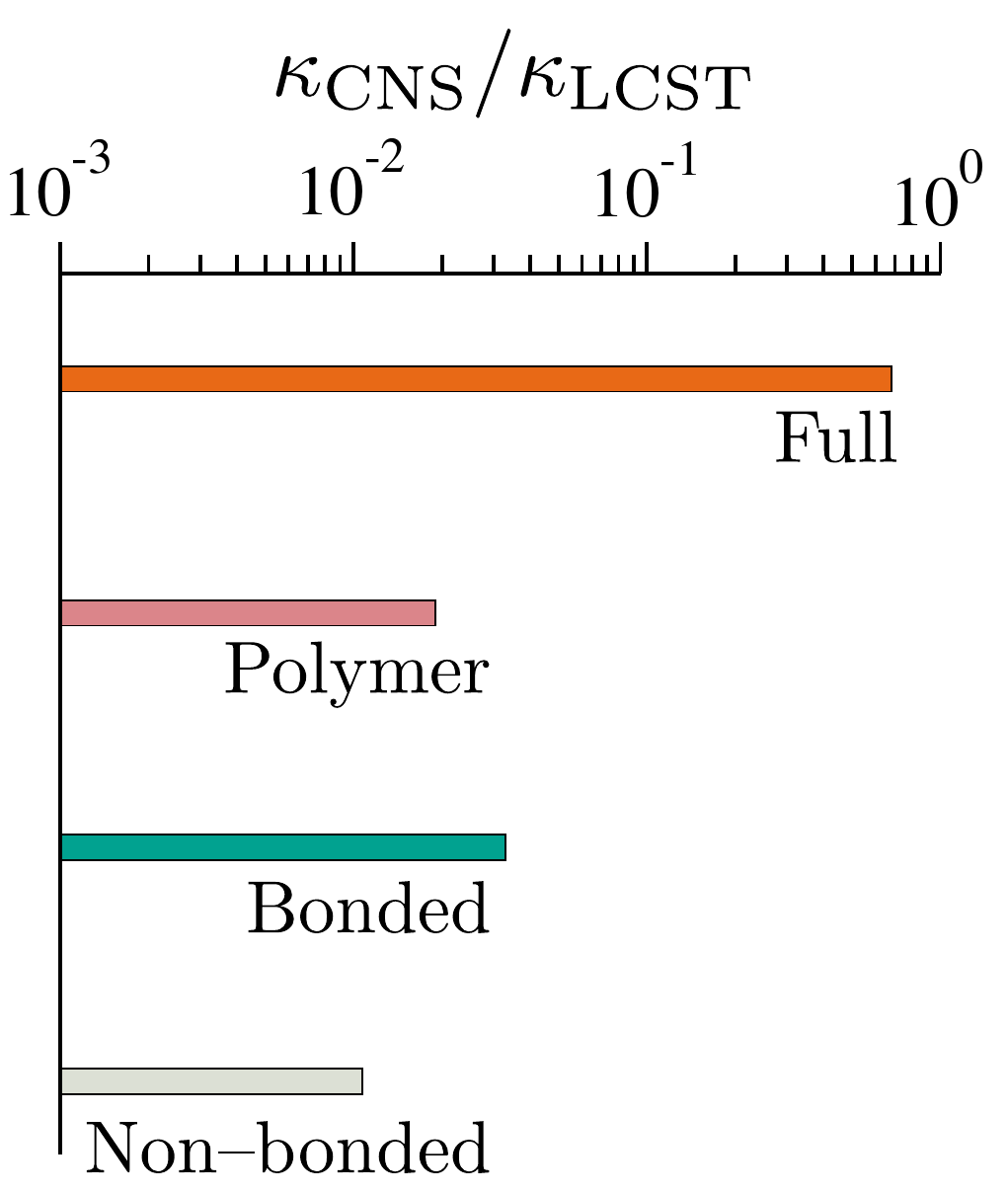}
    \caption{Ratio of the thermal transport coefficients between the two modes of collapse, $\kappa_{\rm CNS}/\kappa_{\rm LCST}$.
    Data is shown for different components contributing the macroscopic $\kappa_{\rm full}$ of the full system. The collapse chain
    for the LCST collapse is taken for a temperature $T 1.15\epsilon/k_{\rm B}$ and for $x_{\rm c} = 0.1$ for the CNS collapse.
\label{fig:Comp}}
\end{figure}

A direct consequence of the above behavior is that-- a globule under CNS remains rather fluffy due to the presence of interstitial cosolvents~\cite{Mukherji14NC,mukherji15jcp}. 
%Such a fluffy (soft) globule contributes to a knocking down of $\kappa_{\rm pol}$. 
Furthermore, the cosolvent mediated (indirect) monomer--monomer contacts provide to additional pathways for energy leakage from the polymer to the cosolvent molecules. A combination of the soft globule/aggregate and the additional energy leakage pathways significantly reduces $\kappa_{\rm full}$. 

To validate the above scenario, we have calculated the temperature relaxation of a polymer $\triangle T = T_{\rm h} - T$ in Fig.~\ref{fig:Tdecay}. For this purpose, a set of non--equilibrium simulations are performed. Here, a polymer is first heated to a temperature $T_{\rm h} = 1.5\epsilon/k_{\rm B}$ under the canonical simulation, while all other particles in a system are kept at $T = 1.0\epsilon/k_{\rm B}$. 
In the following step, $\triangle T$ is allowed to relax under the microcanonical ensemble.
It can appreciated in Fig.~\ref{fig:Tdecay} that $\triangle T$ relaxes about two times faster for $x_{\rm c} = 0.6$ (an expanded chain) than for $x_{\rm c} = 0.1$ (a globule).
This is not surprising given that the increase in $x_{\rm c}$ also increases the number of monomer--cosolvent contatcs and thus the number of heat leakage pathways from a polymer to the bulk solutions. 

Lastly, we have also compiled data showing a ratio of $\kappa$ between a collapse under LCST and CNS--based collapse. It can be appreciated that the contribution from polymer to $\kappa_{\rm full}$ is only about 30--40\% smaller for CNS with respect to LCST, while $\kappa_{\rm pol}$ is only 2--3\% for CNS than in LCST. Therefore, further suggests that the exact chain conformations do not always play an equally important role, instead depend on various other microscopic details. 

%In particular, the $\kappa-$behavior in the polymer solutions is predominantly dictated by the interaction strengths.

\section{Conclusions}
\label{sec:conc}

We have performed molecular dynamics simulations of a generic model to study the heat flow in the smart polymer
solutions across their coil--to--globule transition. For this purpose, we have investigated two different modes of collapse. In one case, we simply use the experimentally investigated LCST polymer collapse in a single solvent~\cite{pnipamLCSTkappa,tian18acsml}. The second case makes use of the well--known phenomenon of polymer collapse in the solvent mixtures~\cite{mukherji13mac,Mukherji14NC}.

Under LCST, $\kappa$ switches exactly at a temperature $T$ where a polymer undergoes its coil--to--globule transition. No such
correlation was observed in $\kappa$ for CNS. These results highlight that the exact polymer conformation plays a lesser important role 
in dictating $\kappa-$behavior in solutions, instead the solvent coordination within the first solvation shell of a polymer dominates the heat flow in polymer solutions. 

Our data for different polymer concentrations $c$ under LCST is consistent with the prior experiments~\cite{pnipamLCSTkappa,tian18acsml} that showed opposite trends in $\kappa$ switching in different $c$ regimes. However, the data under CNS show a new direction that highlights at the need to look beyond the exact polymer conformations. Instead, a more delicate balance between the different solution micro--details dictates heat flow in polymer solutions. It will certainly require more studies to validate our scenario. Moreover, we hope these results will be helpful for the future experimental and simulation studies.\\

\noindent{\bf Acknowledgement:}
This research was undertaken thanks, in part, to the Canada First Research Excellence Fund (CFREF), Quantum Materials and Future Technologies Program. 
Simulations were performed at the Advanced Research Computing Sockeye facility.
D.M. thanks Kurt Kremer and Carlos Marques for very fruitful collaborations on polymer solvation that led to the basics of this work.\\

\noindent{\bf Conflict of interest:} The authors declare no conflicting financial interest.\\

\bibliographystyle{ieeetr}
\bibliography{MukherjiarXiv}

\end{document}